\documentclass[12pt]{article}
\usepackage{newtxtext,newtxmath}
\usepackage{setspace}
\usepackage{authblk}
\usepackage{array}
\usepackage{caption}
\usepackage[a4paper, top=2cm, bottom=3cm, left=3.3cm, right=3.3cm]{geometry}
\usepackage[utf8]{inputenc}
\usepackage[style=apa,backend=biber]{biblatex}
\usepackage{hyperref}
\usepackage{orcidlink}
\usepackage{graphicx}
\usepackage{subcaption}
\usepackage{float}
\usepackage{enumitem}

\usepackage{booktabs}

	\title{\textbf{Forgetful by Design? A Critical Audit of YouTube’s Search API for Academic Research}}
	\author[1]{Bernhard Rieder}
	\author[2,*]{Adrián Padilla}
	\author[3]{Òscar Coromina}
	
	\affil[1]{University of Amsterdam}
	\affil[2]{Euncet Business School}
	\affil[3]{Universitat Autònoma de Barcelona}
	\affil[*]{Corresponding author: Adrián Padilla, apadilla@euncet.com}
	\date{\vspace{-5ex}}

\begin{document}
	\begin{spacing}{1}
				
		\maketitle

	\noindent
\begin{center}
	\hrule
	\vspace{0.5em}
	\small
	\textbf{Acknowledgement}\\ The Version of Record of this manuscript has been published and is available in \textit{Information, Communication \& Society} 2025-11-30 \url{http://www.tandfonline.com/10.1080/1369118X.2025.2591767}.
	\vspace{0.5em}
	\hrule
\end{center}
	
	\vspace{1em} 
	
	\noindent
	\textbf{Citation}\\ Rieder, B.; Padilla, A.; Coromina, C. (2025) Forgetful by design? A critical audit of YouTube’s search API for academic research. \textit{Information, Communication \& Society}, 1-20. \url{http://dx.doi.org/10.1080/1369118X.2025.2591767}.
	
	\vspace{1em}
	
	\noindent

		\begin{abstract}
			This paper critically audits the search endpoint of YouTube’s Data API (v3), a common tool for academic research. Through systematic weekly searches over six months using eleven queries, we identify major limitations regarding completeness, representativeness, consistency, and bias. Our findings reveal substantial differences between ranking parameters like relevance and date in terms of video recall and precision, with relevance often retrieving numerous off-topic videos. We also observe severe temporal decay in video discoverability: the number of retrievable videos for a given period drops dramatically within just 20-60 days of publication, even though these videos remain on the platform. This potentially undermines research designs that rely on systematic data collection. Furthermore, search results lack consistency, with identical queries yielding different video sets over time, compromising replicability. A case study on the European Parliament elections highlights how these issues impact research outcomes. While the paper offers several mitigation strategies, it concludes that the API’s search function, potentially prioritizing ‘freshness’ over comprehensive retrieval, is not adequate for robust academic research, especially concerning Digital Services Act requirements.		
		\end{abstract}
	
		\textbf{ORCID}: 
\begin{itemize}
	\setlength{\itemsep}{0pt}  
	\setlength{\parskip}{0pt}  
	\item[] Bernhard Rieder \url{https://orcid.org/0000-0002-2404-9277}
	\item[] Adrián Padilla \url{https://orcid.org/0000-0001-7668-1322}
	\item[] Òscar Coromina \url{http://orcid.org/0000-0002-6306-4154}
\end{itemize}
\end{spacing}

	\newpage
	\section{Introduction}
		Since the early days of social media, researchers have continually struggled with the question of how to access relevant data for studying these platforms. While the period until roughly 2015 has been characterized as a `Wild West’, where academics could `compile huge troughs of data with few constraints’ \parencite[1582]{puschmannEndWildWest2019}, the aftermath of the Cambridge Analytica scandal contributed to what has been termed an `APIcalypse’ \parencite{brunsAPIcalypseSocialMedia2019}, where platforms like Facebook and Instagram, in particular, heavily curtailed researchers’ capacity to collect data. The following `Post-API Age’ \parencite{freelonComputationalResearchPostAPI2018} saw attempts to fill the void with Industry-Academic partnerships, but the most well-known example, Social Science One \parencite{kingNewModelIndustry2019} shut down after only one round of projects due to delays and unreliable data. The EU’s Digital Services Act (DSA), which came into force in 2022 and continues to be implemented, is set to disrupt the research landscape once more, as it includes specific access provisions for researchers studying `very large online platforms’ \parencite{Leerssen_2024}. In its wake, the major players---including Facebook, Instagram, TikTok, YouTube, and Twitter (now X)---have introduced, repurposed, or reactivated APIs and launched bespoke data access programs specifically for (European) researchers. Although encouraging, these initiatives raise questions about the scope and limitations of access, including coverage, completeness, representativeness, and other `data quality’ issues that affect researchers’ capacity to generate knowledge.
		
		Scholars have long discussed the central, yet ambivalent role APIs play in academic research \parencite[e.g.,][]{gigliettoOpenLaboratoryLimits2012,driscollWorkingBlackBox2014}. Despite the recognition that `social media companies not only control researchers’ access to data, but can also manipulate their systems in a way that affects the findings’ \parencite{grahamBigTechHarming2024} attention to data quality is scarce outside of an overall limited number of studies that focus specifically on the issue. Twitter, long lauded as the platform most open to researchers \parencite[e.g.,][]{trombleWhereHaveAll2021}, is a notable exception as it has received considerable scrutiny when it comes to assessing data quality over the years \parencite{morstatterSampleGoodEnough2013, morstatterWhenItBiased2014, pfefferThisSampleSeems2023}. YouTube, the second most visited website in the world\footnote{\url{https://www.similarweb.com/top-websites/}}, has remained largely unexamined, however. This is particularly surprising, as the YouTube Data API (v3) was already introduced in 2014 and has overall changed relatively little, although the initially very generous daily quotas provided to developers and researchers were steadily reduced over time. When the company introduced its researcher program\footnote{\url{https://research.youtube.com}} in July 2022, it primarily promised eligible academics `as much quota as required for their research’, resulting in `expanded access to global video metadata across the entire public YouTube corpus via API’ \parencite{YouTubeResearchHow}. But even before that, hundreds of studies relying on the API or API-based software like the YouTube Data Tools \parencite{riederYouTubeDataTools2015} had been published in most cases without much scrutiny given to data quality. Certainly, researchers may mention, for example, that the search endpoint---a machine readable interface to YouTube’s search interface---is limited to 500 results per query or that the (now deprecated) `related videos’ feature is only a partial approximation of the platform’s recommender system. Beyond a series of papers discussing specific sampling techniques \parencite[e.g.,][]{zhouCountingYouTubeVideos2011,riederMappingYouTube2020,malikFrameworkCollectingYouTube2017,bartlYouTubeChannelsUploads2018}, we are not aware of any systematic attempts to evaluate the characteristics, quality, and limitations of the data YouTube makes available through its API.
		
		Given that the DSA transforms data access for researchers from a benevolent gesture into a legal obligation, this paper aims to investigate the search endpoint of YouTube’s Data API, which has been frequently used by researchers across various disciplines to analyze the videos published on one of the leading social media platforms. Despite the widespread reliance on the search feature, there is limited understanding of query matching, coverage, representativeness, and the impact of different ranking parameters on the collection of empirical evidence. To address these questions, we developed a strategy to collect the maximum number of videos per query and conducted systematic weekly searches across eleven queries over six months. Our goal was to better understand how videos are made visible (or invisible) through search and what implications this has for the knowledge researchers can generate. In this paper, we provide both a critical analysis and recommendations for addressing the significant problems we encountered.
		
		The paper proceeds in four steps. First, we briefly survey how researchers have been using the search endpoint, introduce existing API audits for social media platforms, and discuss data collection on YouTube. Second, we outline our methodology, including data collection strategies, analytical techniques, and limitations. Third, we present our findings in aggregate statistical terms as well as through a short case study designed to make them more accessible to researchers from less quantitative disciplines. Finally, we discuss implications for research, provide recommendations for future studies using the search endpoint, and suggest how YouTube could amend its systems to better serve academic audiences and comply with legal obligations.
		
	\section{Literature Review}
		\subsection{Research Using The Search Endpoint}
			As one of the most widely used social media platforms, YouTube has attracted considerable attention from researchers across a wide range of disciplines, including sociology, media studies, health communication, and political science. While YouTube’s Data API has several methods that can be used for data collection, including access to channels’ video lists and user comments, the search endpoint, which takes a text query as input and retrieves metadata for up to 500 videos as output, has been the main entry point for researchers. Studies relying on search generally follow one of two methodological strategies.
			
			First, researchers may access the endpoint without any temporal restrictions, either following a more qualitative or exploratory protocol without much concern for questions of completeness or representativeness, or focusing on information visibility, trying to simulate what a user would see if they were to use the same query. For example, Pires et al. \citeyear{piresDeliveryRidersCultural2023} queried the API with names of online delivery services to find videos in wich delivery riders talk about their experiences. Instead of relying on the full 250 videos found, they reduced their sample manually to 40, a strategy often seen in studies that emphasize in-depth qualitative analysis over representative sampling. Research centering information visibility often follows the report by Pandey et al. \citeyear{pandeyYouTubeSourceInformation2010} on H1N1 influenza videos, carrying some variation of `YouTube as a source of information on’ in the title. While these studies sometimes scrape YouTube’s website instead of using the API, they can be considered prototypical for a larger set of work interested in `the type of content average users are likely to come across when searching for information’ \parencite[2]{marchalCoronavirusNewsInformation2020} on what is regularly referred to as the second largest search engine in the world \parencite{khanResearchingYouTubeMethods2022}. Although studies of this kind are affected by questions of completeness and bias, their primary concern from a data collection perspective is the comparability of API- and website-results, given potential factors such as localization or personalization.
			
			Second, and more crucial for the purposes of this paper, numerous studies concentrate on specific events and timeframes, aiming to reconstruct how an issue was portrayed or reported through YouTube videos. Inwood and Zappavigna \citeyear{inwoodConspiracyTheoriesWhite2023}, for example, used the search endpoint to investigate conspiracy videos published in the 24-hr period after the Notre Dame cathedral fire in 2019; Oliva et al. \citeyear{olivaKidsTheseYouTubers2023} tried to reconstruct the debate after Spanish YouTuber El Rubius announced that he was moving to Andorra for tax reasons; Al-Zaman \citeyear{al-zamanSocialMediatizationReligion2022} and Porreca et al. \citeyear{porrecaUsingTextMining2020} used the YouTube Data Tools' 'one-search-per-day' feature---the method we use in this paper and discuss further down---to exhaustively sample, measure, and track videos related to, respectively, Islam and vaccination. These cases underscore the critical need for the search endpoint to provide accurate and consistent results, which is vital for analyzing key social issues in both current studies and future historical research \parencite{wellerDigitalTracesUsergenerated2016}.
			
			Our broader survey of papers employing these methods revealed that researchers pay generally little attention to the search endpoint’s characteristics and implementation details. Many papers do not report the specific day a search was made and the settings for the crucial order parameter, which has striking effects on results, are hardly mentioned. To shed light on what this means for the reliability and validity of YouTube research, particularly when following the second methodological strategy, our paper proposes an exploratory audit of the search endpoint. The following section examines similar research conducted on other social media platforms.
			
		\subsection{Sampling bias and representativeness in APIs}
			Research on social media data has consistently revealed significant challenges related to sampling bias and representativeness, even when data is collected through sanctioned platform APIs. This has raised broader questions about the integrity of research relying on APIs that function as `black boxes’ primarily designed to serve business or operational needs \parencite{driscollWorkingBlackBox2014}. However, given that academics often lack viable alternatives, it is essential to understand the biases that exist and how they impact research.
			
			A foundational contribution to this context comes from Morstatter et al. \citeyear{morstatterSampleGoodEnough2013}, who compared Twitter’s Streaming API, which provides a 1\% sample of public data, with the platform’s Firehose API, granting full access. Their findings highlight significant discrepancies between the two, including variances in geographic coverage, network structure, and topic distribution, particularly pronounced during high-activity events. Further investigations by Morstatter et al. \citeyear{morstatterWhenItBiased2014} and Tromble et al. \citeyear{trombleWeDontKnow2017} delved into the conditions under which such biases emerge, pointing to real-time trends and data spikes during major events as key amplifiers. Together, these studies underscore the importance of caution when interpreting findings derived from API-restricted data, as these samples may not adequately represent the broader platform activity.
			
			Similarly, Rieder et al. \citeyear{riederDataCritiqueAnalytical2015} provided a critique of Facebook’s Graph API, emphasizing the limitations of data access in the wake of increasing restrictions. They examined a large dataset obtained from Facebook Pages and identified discrepancies in the representativeness of user interactions, pointing out how platform design influences what researchers can study. Villegas \citeyear{villegasFacebookItsDisappearing2016} further explored challenges in data retention on Facebook, documenting how content such as posts and comments can `disappear’ over time due to platform-specific data curation practices. Ho \citeyear{hoHowBiasedSample2020} extended this discussion by reverse-engineering Facebook’s ranking algorithms, revealing how specific types of content are prioritized or excluded from API results. These studies highlight that, regardless of the platform, APIs may impose unseen biases that can affect research outcomes by limiting researchers’ access to high-quality samples.
			
			Even when platforms offer expanded access under academic research programs, significant challenges may persist. Pearson et al. \citeyear{pearsonMarginErrorSystematic2024} conducted a systematic audit of TikTok’s research API---launched in July 2022---by comparing API-retrieved metrics (e.g., comments, likes, and views) with the data displayed on the site’s user interface. Their findings revealed systematic discrepancies, again putting into question the utility and reliability of official data access provisions.
			
			While the idiosyncrasies of these social media platforms are increasingly well-documented, few studies have specifically addressed data collection on YouTube. Given the platform's central role in the digital media ecosystem, the recent expansion of API access for eligible researchers, and the widespread reliance on the search endpoint, it is crucial to better understand the level of data quality one can expect.
		\subsection{The YouTube Data API}
			Although YouTube has not been the primary focus in the ongoing discourse surrounding the `APIcalypse’ \parencite{brunsAPIcalypseSocialMedia2019}, the platform’s data access provisions have undergone two significant changes since the introduction of the YouTube Data API (v3), both of which have directly affected researchers. First, API quotas have been progressively reduced, making large-scale studies increasingly difficult without resorting to alternative methods such as web scraping. While YouTube’s researcher program offers some relief by providing vetted projects with enhanced access, its scope remains limited and subject to the platform’s vetting process. Second, the removal of the `related videos’ feature in 2023---a feature that, although limited, allowed some examination of algorithmic pathways---has further constrained researchers’ ability to analyze YouTube’s recommendation system.
			
			The quality of data access through YouTube’s Data API presents a mixed picture. On the one hand, it offers a variety of methods to collect detailed data, with generally fewer restrictions than other platforms. On the other hand, there are limitations that directly affect researchers studying content dynamics or monetization, such as the unavailability of video transcripts, the absence of a flag to identify Shorts, the impossibility to know whether a video is monetized, or the absence of reasons for why a video or channel is missing. The official API documentation is often silent when it comes to reporting or explaining these and other limitations. Developers therefore exchange technical insights, undocumented behavior, and workarounds - such as how to check whether a video is a Short - on forums like Stack Overflow\footnote{\url{ https://stackoverflow.com/questions/71192605/how-do-i-get-youtube-shorts-from-youtube-api-data-v3}}. But these solutions often involve scraping the necessary information from the platform’s web interface, a practice that potentially forces researchers into a legal grey zone.
			
			One of the areas where researchers have addressed data collection specifics on YouTube are methods for collecting representative samples from the company’s vast content library. The most feasible method to create a random sample for all of YouTube, `random prefix sampling’, was initially introduced by Zhou et al. \citeyear{zhouCountingYouTubeVideos2011} and more recently validated by McGrady et al. \citeyear{mcgradyDialingVideosRandom2023}. Bärtl \citeyear{bartlYouTubeChannelsUploads2018} experimented with random search queries to trace overall trends on YouTube over time. While these methods are effective for analyzing macro-level trends, they are hardly practical for topic-specific studies as a random sample would have to be exceedingly large to contain enough videos covering individual cases or issues. The same goes for the large-scale crawling method implemented by Rieder et al. \citeyear{riederMappingYouTube2020}. Most researchers studying specific issues therefore rely on the Data API’s search function to create a collection of videos to analyze.
			
			As Malik and Tian \citeyear{malikFrameworkCollectingYouTube2017} noted, however, the search endpoint’s restriction to 500 results makes it unsuitable for large-scale data retrieval. For many studies this limitation may not be a barrier, particularly for those focused on less popular topics, specific geographic and linguistic contexts, or narrowly defined timeframes, where the number of relevant videos may be smaller. However, questions remain on how YouTube’s search system actually selects and ranks videos. While we know little in terms of technical specifics, Rieder et al. \citeyear{riederRankingAlgorithmsRanking2018} demonstrated that search rankings can fluctuate significantly, especially for `newsy’ queries that experience high levels of uploads and user engagement. This volatility underscores the importance of timing in data collection and highlights the challenges of replicating studies that depend on search-based data. Finally, Efstratiou’s (\citeyear{efstratiou_youtube_2025}) work corroborates previous findings while revealing new insights into the substantial variability and inconsistency characterizing content retrieval via the YouTube API. His results suggest that search outputs exhibit randomization patterns, apparently influenced by popularity metrics and temporal factors.

			Given the prevalence of query-based approaches in YouTube research, this paper explores several critical questions surrounding the search endpoint: How do different ranking parameters change which videos are matched to a given query? How does temporal distance affect search outcomes? What are the implications of these findings for researchers? And finally, how can researchers adapt their data collection strategies to mitigate limitations and enhance the robustness of their studies?

	\section{Method}
		To analyze the Data API’s search endpoint in detail, this paper takes a `maximalist’ approach and tries to collect exhaustive samples for a set of queries. We rely on a method used by researchers 
		\parencite[e.g.,][]{porrecaUsingTextMining2020,al-zamanSocialMediatizationReligion2022,violotShortsVsRegular2024} to circumvent the search endpoints’ limit to 500 results per query. Instead of making only one call to the search endpoint, the 'one-search-per-day' principle cuts a longer timeframe into several days (or other intervals) and makes a separate call for videos published on each one of these days. For example, when searching for videos published in a specific week, one could make seven individual calls to the API, each one limited to a single day, yielding up to 3500 videos instead of 500. If the search endpoint were to allow for full coverage of the underlying population of relevant videos, this would enable full retrieval for all but the largest topics, where video production exceeds 500 videos per day. For this paper, we used the YouTube Data Tools \parencite{riederYouTubeDataTools2015}, which implemented a `one search per day’ feature in 2018, and ran searches for eleven queries weekly over the span of six months starting in April 2024. For each query, we set the starting date to October 15, 2023 and requested up to 500 results per day.
		
		The Data API’s search endpoint, however, adds another layer of complexity by offering several ranking principles through the order parameter, with limited information on how they differ beyond very general terms. While YouTube mentions\footnote{\url{https://www.youtube.com/intl/en_us/howyoutubeworks/product-features/search/}}
		 content match, engagement, and quality as the main factors for search overall, the API documentation on the differences between ranking principles is sparse:
		
		\begin{itemize}
			\item \textbf{\texttt{date}} – Resources are sorted in reverse chronological order based on the date they were created.
			\item \textbf{\texttt{rating}} – Resources are sorted from highest to lowest rating.
			\item \textbf{\texttt{relevance}} – Resources are sorted based on their relevance to the search query. This is the default value for this parameter.
			\item \textbf{\texttt{title}} – Resources are sorted alphabetically by title.
			\item \textbf{\texttt{viewCount}} – Resources are sorted from highest to lowest number of views. For live broadcasts, videos are sorted by number of concurrent viewers while the broadcasts are ongoing.
		\end{itemize}

		Since relevance is the default option for the order parameter and most papers\footnote{Although we did not systematically survey papers using the search endpoint, we reviewed a substantial number and observed that few studies report on the ranking setting used.} seem to rely on it, we focused on this ranking principle. To keep the already very high API quota cost for our project in check, we only used date ranking for comparison after we noticed that other parameters were producing largely similar results. We chose our queries based on our issue expertise and to roughly represent `typical’ keywords researchers may use, covering health, political, and popular culture topics, some connected to a specific event. While this limited selection reduces generalizability, throwing a larger net was infeasible due to the high API quota cost of the `one search per day’ approach. The queries we chose were: `Andrew Tate’, `Angela Merkel’, `Carles Puigdemont’, `European Parliament election’, `Eurovision’, `Gaza ceasefire’, `Geert Wilders’, `Mukbang’, `Obesity’, `Ukraine war’, and `Ursula von der Leyen’. We added quotation marks to each query; however, to enhance readability, we use square brackets to mark queries in the following sections.
		
		To analyze the collected search results, we pursued two different yet complementary approaches. In the first step, we identified broader patterns across queries and documented our findings primarily in quantitative terms. Our initial examination of the different ranking principles yielded varied results, with the relevance setting providing substantially more videos, many of which seemed unrelated to the search queries. Therefore, we first investigated content matching by searching for the query text in the title, description, and tags of each video. Next, we assessed temporal coverage, as we had reasons to believe that the search endpoint exhibits a strong recency bias, making historical research problematic. Finally, we examined sample consistency to understand how results differed between weekly searches. Although our analysis was limited to eleven queries, which somewhat constrains generalization, these three approaches yielded a mostly coherent picture that likely applies to the search feature overall. In the second step, we focused on a specific query---[European Parliament election]---to incorporate qualitative elements and contextualize the consequences of our findings, making them more accessible to a less quantitatively inclined audience.
		
	\section{Findings}
		\subsection{Query precision: important differences between ranking parameters}
			Our first analysis focuses on the differences between the relevance and date ranking principles. Table~\ref{tab:Table_1} summarizes our findings and indicates, depending on the query, small to very large differences in terms of the number of videos retrieved. While some of the highest volume queries like `Andrew Tate', `Eurovision', `Mukbang', and `Ukraine war' produced `only' up to twice the number of videos for relevance, this went up to over three (`Obesity'), four (`Gaza ceasefire'), five (`Angela Merkel') or eight (`European Parliament election') times for others. Since some of the lower volume queries (`Carles Puigdemont', `Geert Wilders', `Ursula von der Leyen') did not yield such large differences,  this is not simply an effect of YouTube’s relevance ranking trying to fetch more results for less prominent topics.

\begin{table}[ht]
	\centering
	\scriptsize
	\resizebox{\textwidth}{!}{%
		\begin{tabular}{p{3.2cm}rrrrr}
			\toprule
			\textbf{Query} & \multicolumn{2}{c}{\textbf{Precision (\%)}} & \multicolumn{2}{c}{\textbf{Videos Retrieved}\footnotemark} & \textbf{\% Volume Diff.} \\
			& \textbf{Date} & \textbf{Relevance} & \textbf{Date} & \textbf{Relevance} & date vs relevance \\
			\midrule
			Andrew Tate & 96.6 & 91.9 & 114469 (110522) & 135459 (124448) & +18.3\% (+12.6\%) \\
			Angela Merkel & 82.5 & 33.6 & 4629 (3821) & 27433 (9229) & +492.6\% (+141.5\%) \\
			Carles Puigdemont & 98.8 & 93.9 & 7311 (7223) & 11926 (11194) & +63.1\% (+54.9\%) \\
			European Parliament\\Election & 57.6 & 17.2 & 8933 (5145) & 72767 (12499) & +714.5\% (+142.9\%) \\
			Eurovision & 95.3 & 53.8 & 72580 (69201) & 141927 (76383) & +95.5\% (+10.3\%) \\
			Gaza Ceasefire & 84.5 & 50.6 & 23501 (19859) & 107329 (54290) & +356.7\% (+173.3\%) \\
			Geert Wilders & 95.8 & 68.3 & 5089 (4874) & 9663 (6603) & +89.8\% (+35.4\%) \\
			Mukbang & 95.3 & 82.3 & 226394 (215813) & 254098 (209005) & +12.2\% (-3.1\%) \\
			Obesity & 77.3 & 33.1 & 44737 (34593) & 157398 (52124) & +251.8\% (+50.6\%) \\
			Ukraine War & 84.0 & 73.4 & 123028 (103290) & 159649 (117196) & +29.7\% (+13.4\%) \\
			Ursula von der Leyen & 90.8 & 78.0 & 14805 (13440) & 25709 (20049) & +73.6\% (+49.1\%) \\
			\bottomrule
		\end{tabular}
	}
	\caption{Dataset overview for our eleven queries.}
	\label{tab:Table_1}
\end{table}
\footnotetext{The numbers in parentheses indicate search results after applying keyword filtering.}

			Since manual inspection, especially for [European Parliament election] (see section 5), showed that many of the retrieved videos were unrelated to the query, we operationalized a measure of precision by searching for our queries\footnote{We modified some of the queries for better results, for example, for the query [european parliament election], we searched for 'europe* AND elect*' and we only used last names for politicians.} in the title, description, and tags of each video. 
			
			While this method can be problematic from a research perspective due to keyword spam or keywords only appearing in the videos themselves, it allows for good comparison between the two ranking principles under scrutiny. The results (Table~\ref{tab:Table_1}) show higher precision for date in all instances, but the difference again varies between queries. We observed lower differences for exceptionally high-volume queries like `Andrew Tate', `Mukbang', and `Ukraine war' and could speculate that such high-volume queries provide enough videos to saturate relevance ranking; but given that YouTube’s search system almost certainly\footnote{Google announced in 2019 that they were using BERT for search https://blog.google/products/search/search-language-understanding-bert/} uses a large language model like BERT \parencite{devlinBERTPretrainingDeep2019} to represent text, we may see the effects of semantic shifts that are specific to each query. 
			
			Although the higher level of what information retrieval researchers \parencite[eg.,][5]{manningIntroductionInformationRetrieval2008} call precision (which percentage of videos match the query) may render date search more attractive at first glance, relevance has in almost all cases higher recall (how many matching videos are retrieved), yielding more videos that match the search terms in the title, description, or tag field. The exception is [Mukbang], the highest volume query in our sample, where the 500 results per day limit probably keeps relevance ranking from finding even more videos, conferring an advantage to the higher-precision date matching.
			
			While one could argue that the `expansive’ behavior of relevance option simulates what users see when they search, as relevance is also the default option on the website and app, it renders quantitative overviews for topics problematic. As we will argue further down, researchers may want to use relevance ordering in combination with keyword filtering for all but the largest volume queries.


		\subsection{Temporal distance: videos can no longer be found}
\begin{figure}[H]
			\centering
			\begin{subfigure}{\textwidth}
				\centering
				\includegraphics[width=\textwidth, height=0.35\textheight, keepaspectratio]{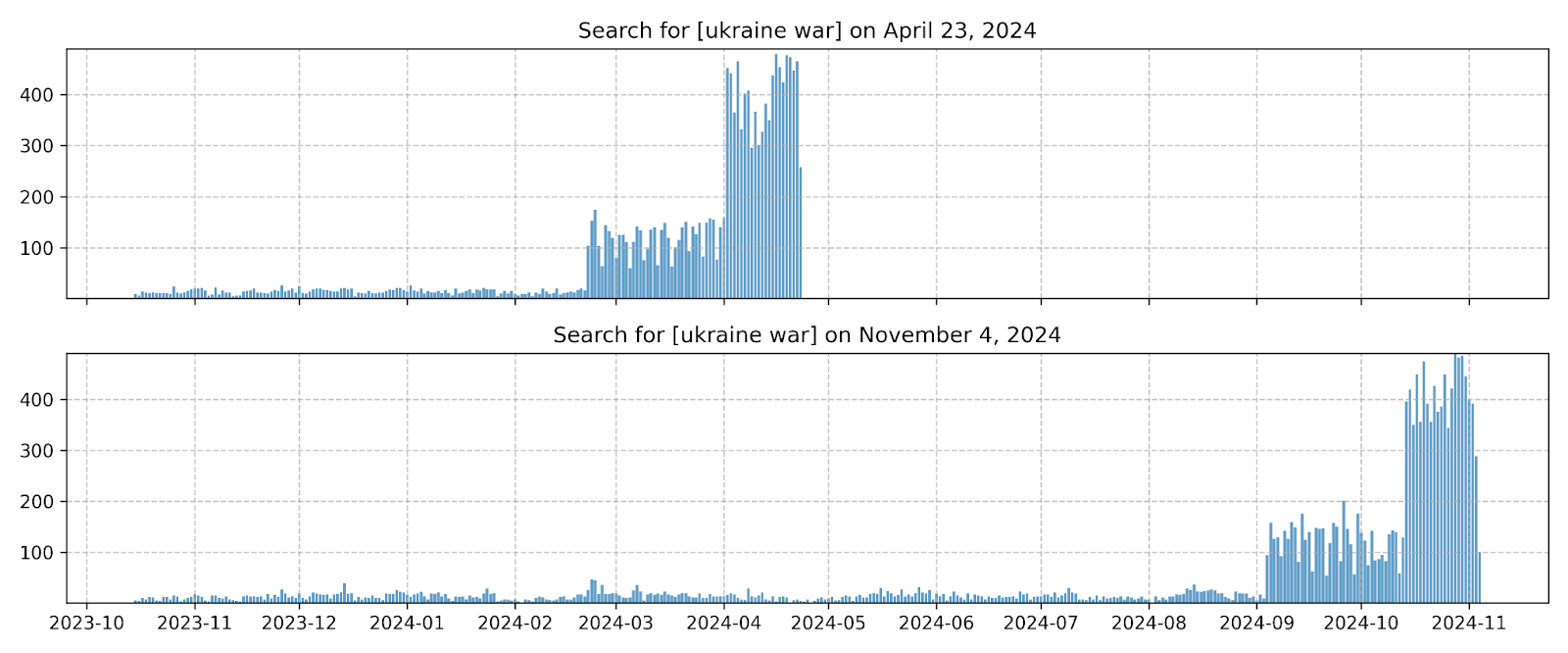}
			\end{subfigure}
			\vspace{0.5cm}
			\begin{subfigure}{\textwidth}
				\centering
				\includegraphics[width=\textwidth, height=0.35\textheight, keepaspectratio]{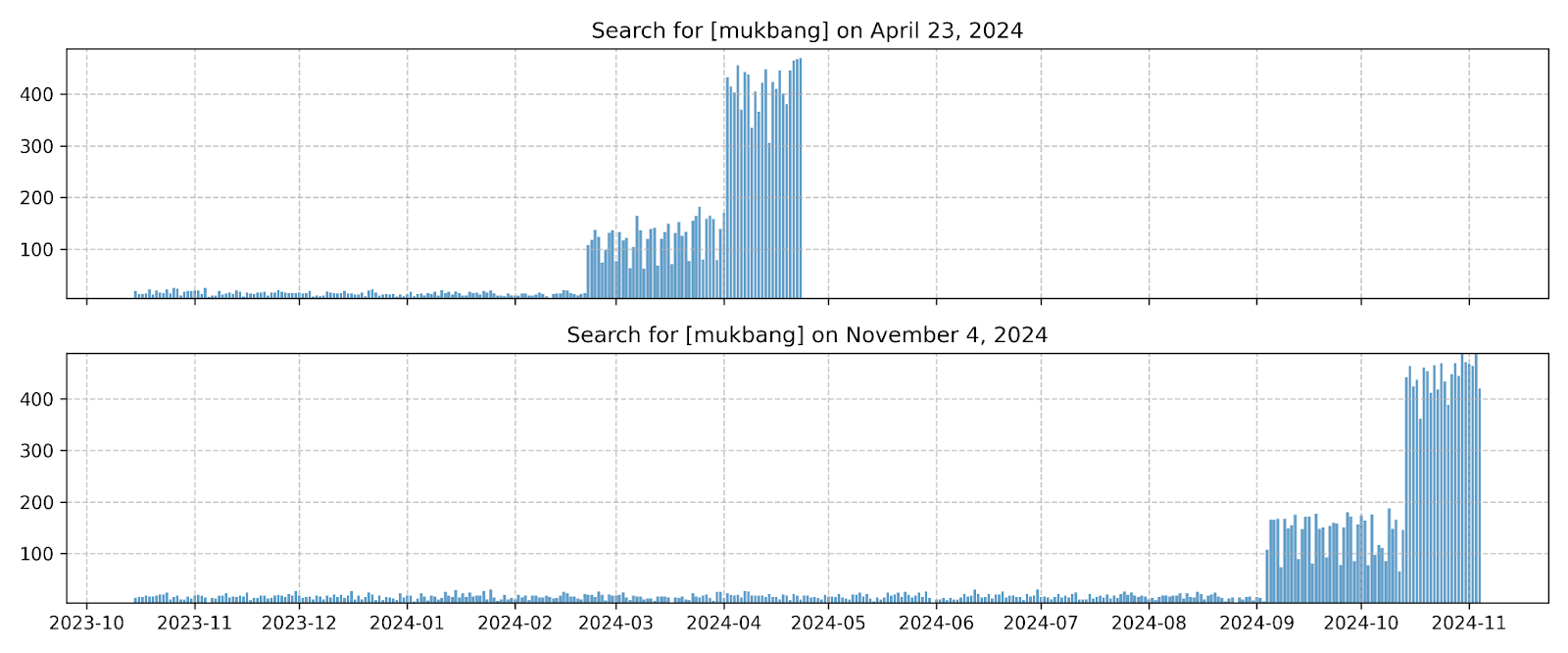}
			\end{subfigure}
			\vspace{0.5cm}
			
			\caption{Top two: Videos per day, published from October 15, 2023 onwards, for the queries `Ukraine war' (top) and `Mukbang' (middle), searched on April 23, 2024 and November 4, 2024.}
			\label{fig:figure_1}
		\end{figure}

		Although researchers must consider the important differences between ranking parameters, this paper’s main findings concern temporal distance. The following charts summarize the central problem with YouTube’s search function emblematically:

		Figure~\ref{fig:figure_1} shows very similar---and very problematic---behavior for two queries, `Ukraine war' and `Mukbang', that should have little to nothing in common. We searched for both queries on separate dates (April 23, 2024 and November 4, 2024), using the same starting point (October 15, 2023). In both cases, we notice a period of 20 days before the search date where results average around 450 videos per day, followed by 40 days with around 130 videos per day, ending in a flat tail with about 20 videos or less found per day. A large majority of videos found in the first search (April) no longer appeared in the second search (November), although most, if not all, were still available on the platform. The three ``phases''---head, middle, and tail---could be distinguished in most of our queries (Figure 2), regardless of when the search was conducted. This means that the volume of videos retrieved does not reflect the actual content published on the platform but is instead shaped by design decisions that result in a very significant recency bias. For researchers studying a particular event or timeframe, this means that the number of videos they can retrieve will drop heavily only 20 days after the chosen date and again 40 days later. Scholars that miss these cutoffs, which includes anybody doing any kind of historical research, should expect to retrieve markedly impoverished samples.

		\begin{figure}[H]
			\centering
			\includegraphics[width=\textwidth, height=0.35\textheight, keepaspectratio]{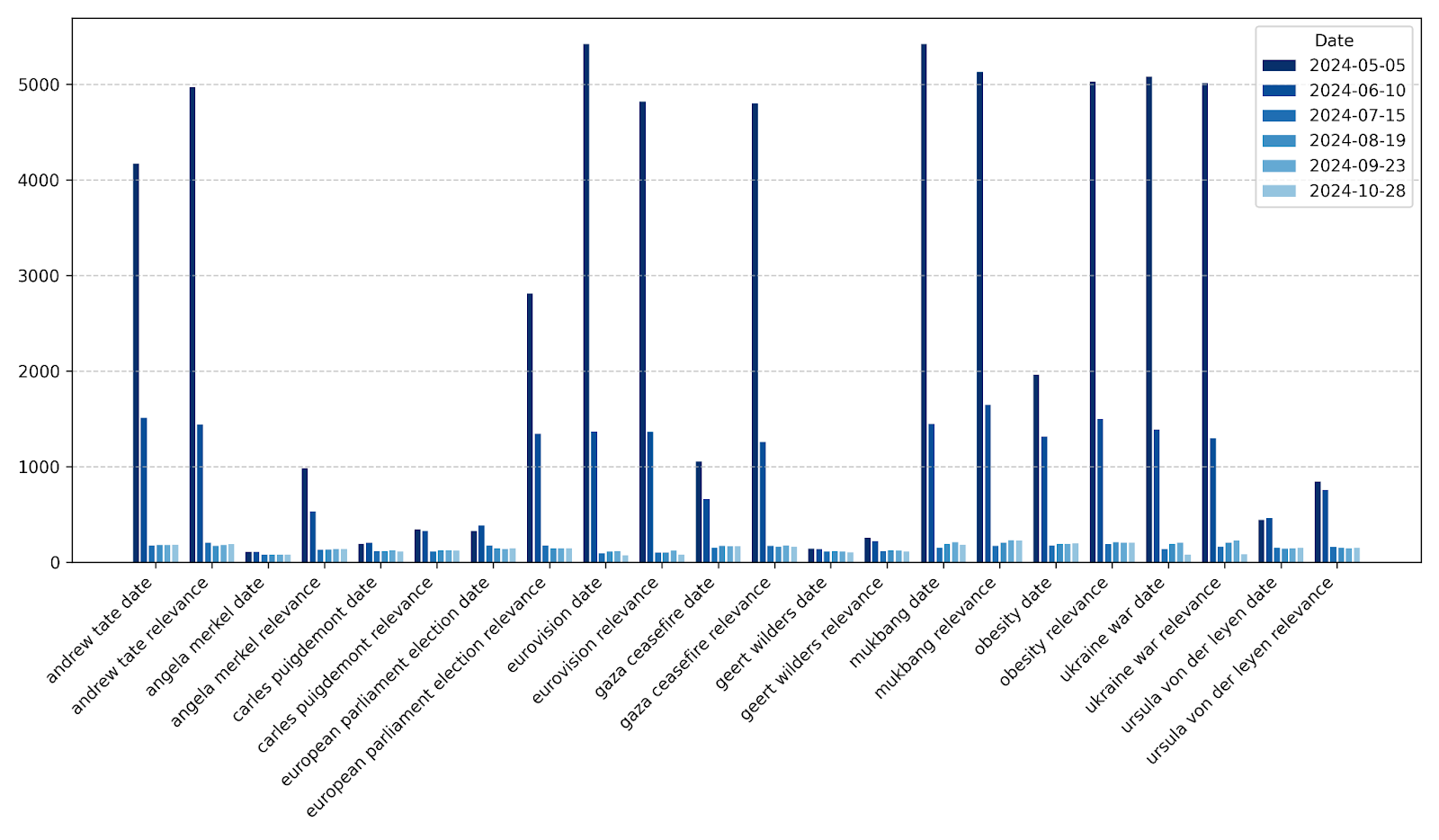}
			\caption{Number of videos published between April 20, 2024 and May 4, 2024 for all queries, searched on six separate dates with both relevance and date ranking.}
			\label{fig:figure_3}
		\end{figure}
			As Figure 2 shows, the eﬀect is more pronounced for higher volume queries and for relevance ranking. Searches ranked by date for [Angela Merkel], [Carles Puigdemont], and [Geert Wilders], show little or no drop-oﬀ at later dates, although we can observe such an effect when using the default relevance settings.
			
			A plain explanation for this behavior is that the search endpoint is not designed with researchers and their speciﬁc needs in mind. While we have not been able to ﬁnd any kind of public commentary on the matter, earlier publications coming out of YouTube itself \parencite[e.g.,][]{covingtonDeepNeuralNetworks2016} emphasize concepts such as 'freshness' and put a strong emphasis on serving recent videos. The impression that YouTube’s search feature is indeed more of a recommendation engine than a traditional information retrieval system was further strengthened when we applied keyword ﬁltering to measure precision for the search phases we discovered. Except for [Mukbang],where the head contained the highest percentage of matching videos, all our queries showed a systematic gain in precision when moving from the head to the tail. The overall emphasis of relevance ranking, in particular, is serving recent videos that may only be peripherally related to the actual search query.

		\subsection{Consistency: datasets cannot be replicated}
			A third issue arising from the behavior of YouTube’s Data API is the uncertainty regarding the consistency of samples collected through the search endpoint. Our analysis indicates that the videos retrieved in each extraction may vary, even when using the same parameters.

			To test for consistency, we compared the results for three separate searches (Figure~\ref{fig:figure_3}), each one a week apart, over the same timeframe. We searched for videos published the week before the first search to make sure that the timeframe would still be in the 20-day head phase for each search. We then separated `new' and `old' videos depending on whether a video was present in the week or weeks before. For example, using relevance, `Mukbang' yielded 2407 videos for the first week, all considered new, since this was our first search. The week after, we received only 775 videos from the first search and 1510 that were new. In the last week, we still added 1437 new videos published in the same timeframe. These results indicate that the number of videos retrieved by the API not only fluctuates across extractions but also that each search introduces new videos not seen before and omits videos that were present in previous iterations.
			
			Unsurprisingly, this effect is more pronounced for high-volume queries like `Mukbang', `Obesity', `Eurovision', and `Ukraine war'. But even lower volume queries like `Carles Puigdemont' and `Angela Merkel' are affected, indicating that this is not just due to result volume hitting some limitation. We also observe that relevance ranking is less consistent than date ranking, although the latter is still far from stable. This lack of consistency hinders the ability to replicate studies, potentially leading to divergent conclusions even when data collection and analysis are conducted under identical conditions.
		
			\begin{figure}[H]
			\centering
			\includegraphics[height=0.95\textheight, keepaspectratio]{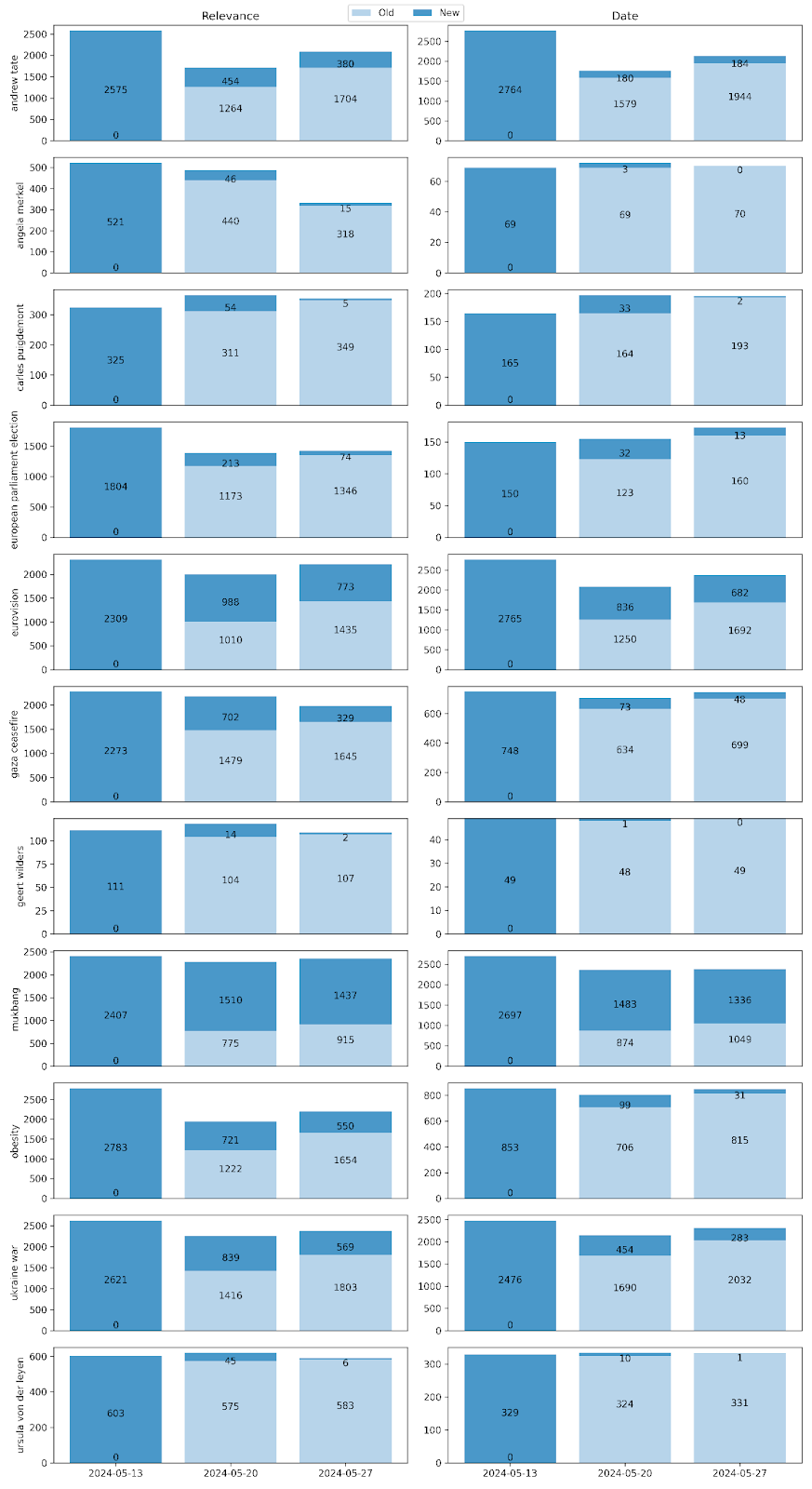}
			
			\caption{New and old videos published between May 7 and May 13, 2024 for searches over three consecutive weeks.}
			\label{fig:figure_3}
			\end{figure}
\newpage
	\section{Case study: the European Parliament elections}
			To make these issues more tangible, especially for audiences outside of technical disciplines such as the computational social sciences, this section delves into the effects of the observed limitations through a case study focused on the European Parliament elections held between June 6 and 9, 2024. We posit that scholars interested in studying the election campaign would likely select the month leading up to June 9 as their observation period and default to relevance ranking. For data collection, we chose five search dates, each spaced five weeks apart, beginning on June 10.

			When looking at the data, we first notice, in line with Table 1, that the precision for this query, again assessed through keyword matching, is particularly low (57.6\% using date and 17.2\% with relevance as ranking method).  A significant number of videos, for example, deal with UK immigration issues and the general elections in India that happened during a similar timeframe. In these situations, where many of the retrieved videos are unrelated to our case study, it is necessary to consider applying keyword filtering. In our example, filtering with the terms `europe* AND elect*' (which includes variations such as `european', `election', `electoral', etc.) in title, description, or tags, the combined dataset of 8941 videos is reduced to only 2244 (Table~\ref{tab:table_2}). Despite our initial query `European Parliament election' explicitly including quotes, relevance throws a much wider net and even date---although much more precise---includes many videos that can hardly be considered on topic. Researchers will have to decide within the context of their research projects whether keyword filtering is required or appropriate, but at least for this case study, the combination of relevance ranking and keyword filtering yields the best results.

\begin{table}[ht]
	\centering
	\resizebox{\textwidth}{!}{%
		\begin{tabular}{p{2.2cm}rrrrr}
			\toprule
			\textbf{Search date} &
			\multicolumn{2}{l}{\textbf{Relevance}} &
			\multicolumn{2}{l}{\textbf{Date}} \\
			& \begin{tabular}{@{}r@{}}No. of videos\\(\% loss)\end{tabular} &
			\begin{tabular}{@{}r@{}}Filtering keywords\\(\% loss)\end{tabular} &
			\begin{tabular}{@{}r@{}}No. of videos\\(\% loss)\end{tabular} &
			\begin{tabular}{@{}r@{}}Filtering keywords\\(\% loss)\end{tabular} \\
			\midrule
			2024-06-10 & 8354 & 2105 & 2259 & 1458 \\
			2024-07-15 & 3035 (-63.6\%) & 1246 (-40.8\%) & 1498 (-33.6\%) & 1004 (-31.1\%) \\
			2024-08-19 & 635 (-92.3\%) & 500 (-76.2\%) & 586 (-74.0\%) & 480 (-67.0\%) \\
			2024-09-23 & 621 (-92.5\%) & 491 (-76.6\%) & 593 (-76.7\%) & 481 (-67.0\%) \\
			2024-10-28 & 610 (-92.6\%) & 485 (-76.9\%) & 580 (-76.3\%) & 472 (-67.6\%) \\
			\midrule
			\textbf{Overall} & 8941 & 2244 & 2383 & 1501 \\
			\bottomrule
		\end{tabular}
	} 
	\caption{Number of retrieved videos published between May 9 and June 9, 2024 for the query `European Parliament election' on five search dates.}
	\label{tab:table_2}
\end{table}

			Table~\ref{tab:table_2} shows the number of videos we found on our five consecutive search dates, spaced five weeks apart, starting on June 10, the day after the elections. The extraction timeframe for the five searches is identical: the month leading up to June 9. This experiment helps us explain how the same extraction, conducted at different times, can yield different results and observe how the number of videos the API offers for this period decreases over time. We again notice that \texttt{relevance} retrieves not only many more videos, but also more videos that match our keywords. 
			
			Most importantly, however, we notice a very significant drop-off in volume as we move away from the observation period: compared to the first search on June 10, waiting five weeks means that results are reduced by about 41\% with keyword filtering (from 2105 to 1246) and 64\% without (from 8354 to 3035). If we search ten weeks later, we lose 76\% and 92\% respectively, compared to the search directly following the end of the observation period. After that, search volume remains relatively stable. 
			
			This lines up with the head, middle, and tail sections identified further up; and while date ranking shows less precipitous drop-offs, the same pattern applies. Finally, since the overall number of videos found across all five searches is higher than the number of videos in the first search, we conclude that this is not merely a drop-off problem but also a consistency issue.
			
			To better understand these findings, we examined potential differences between missing and available videos, starting from the intuition that videos with higher view counts or engagement metrics would be more likely to remain in search results. This, however, was not the case. It is conceivable that metrics such as click-through rates---reflecting what users actually watch after a search---play a role, but these data are not accessible to researchers. While future work may shed more light on these dynamics, definitive causal explanations are unlikely without internal platform access.

			\begin{figure}[H]
				\centering
				\includegraphics[width=1\textwidth]{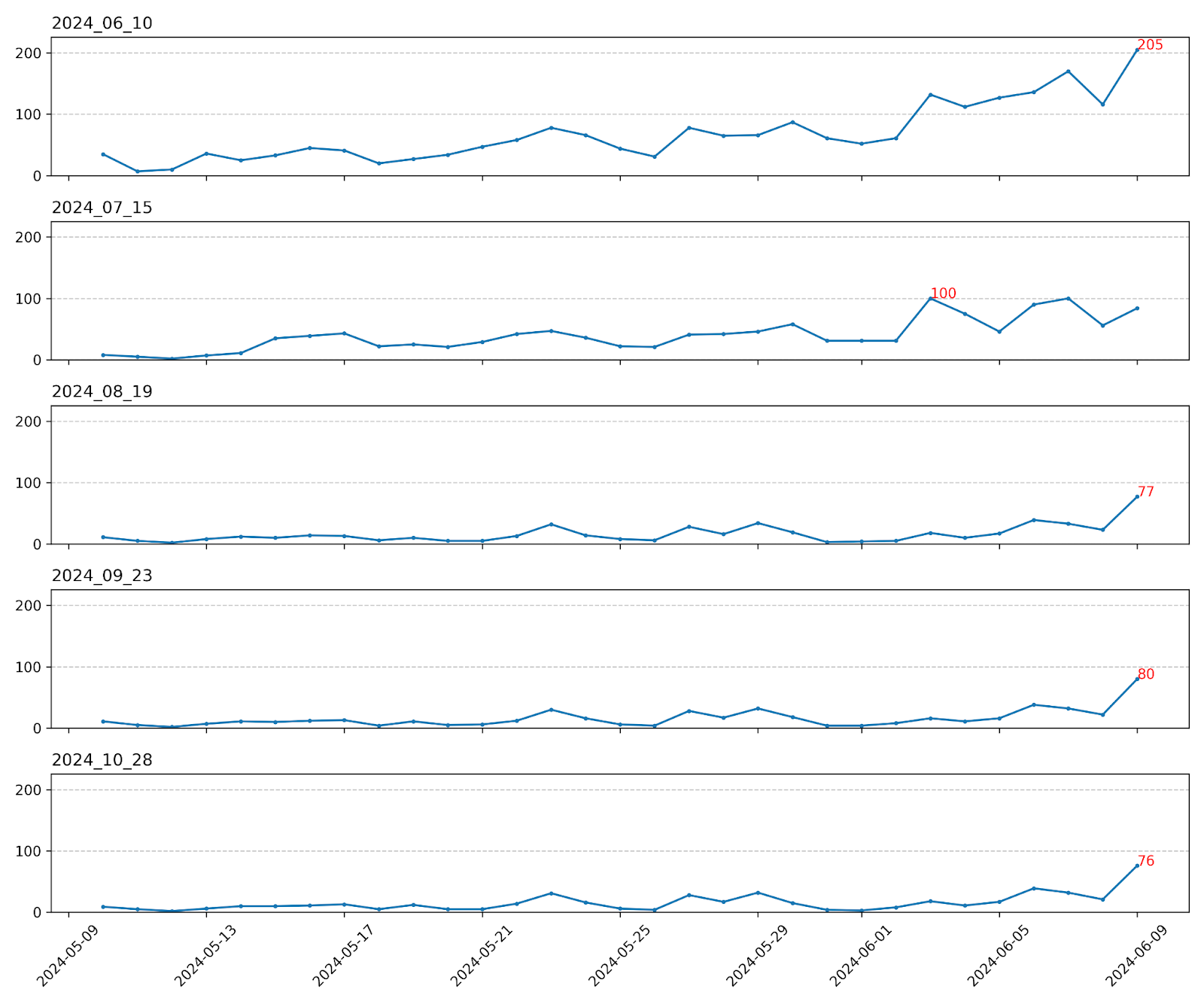}
				\caption{Videos per day for five searches over the same observation period, with keyword filtering.}
				\label{fig:figure_4}
			\end{figure}
			
			How does the loss of videos occur throughout the extractions, and how would this affect our understanding of a phenomenon's evolution? As Figure~\ref{fig:figure_4} illustrates, the reduction in findable videos does not completely flatten temporal variation, as the general form of the curve remains at least somewhat similar (e.g., an uptick towards the end of the period). However, the differences are still significant. In our case, we believe that evaluating the evolution of the European election campaign based on any search other than the first can lead to a biased assessment of peak moments or the overall volume of the conversation on YouTube. While we can never be certain that we have retrieved all relevant videos, quantitative claims about the volume of videos published around an issue become increasingly problematic as we move further in time from the period of interest.
			
			Since researchers often focus on the most viewed videos when selecting units to analyze, we also investigated how many of the top 10 videos on a certain search date are still present in the weeks that follow (whether in the top 10 or elsewhere in the dataset). We again found that increased temporal distance implies a significant loss, at least for this case study. For example, of the top 10 most viewed videos in the first of our five searches, nine were still present five weeks later, dropping down to five afterwards. This means that over half of the videos that had important view numbers just after the end of the European Parliament elections were no longer found four months later, although all of them were still available on YouTube when we checked manually. 
			
	\section{Discussion and Recommendations}
		Our findings document three main problems with the search endpoint of YouTube’s Data API. First, ranking parameters exhibit very different behaviors, with relevance returning the largest number of videos but also---depending on the query---many videos that are clearly unrelated to the issue specified by the search query. Second, we observed a consistent and very significant drop-off in search coverage as search dates moved further away from video publication dates. Researchers hoping to study an event through the lens of YouTube videos will find strikingly fewer videos if they wait with their search, particularly if they exceed 60 days, even though these videos remain on the platform. As our European Parliament election case study has shown, the videos no longer covered are not low-visibility clips, but potentially some of the most viewed, making the loss even more problematic. Third, the API does not provide a stable set of videos for a given query, meaning that identical queries for the same topic can yield different outputs over time.
		
		While our study design is constrained by the number and representativeness of the chosen queries, the six-month observation period, and the inability to examine potentially salient content features such as transcripts, these limitations do not affect the central conclusion that the search endpoint exhibits clear and systematic problems. 
		
		Although these problems manifest to varying degrees for different issues, with high-volume queries likely being the most problematic, they introduce significant uncertainty regarding completeness, representativeness, consistency, and bias in any research project relying on the search endpoint. Research that investigates what YouTube serves about a particular subject to its users at a specific point in time may be less affected. However, any project aiming to study the videos published around a given topic can hardly be confident about their sample. While metadata for any public video can still be collected, if videos no longer appear in searches after a few weeks, it becomes impossible to reconstruct how an issue was portrayed or reported through YouTube videos. Efforts to quantify the spread of misinformation during a public health crisis, identify the first outlet to report a major event, or trace the evolution of political discourse over time are significantly constrained. Given YouTube’s importance in the larger digital media environment, these findings contribute to existing critiques of the use of APIs in academic research \parencite[e.g.,][]{gigliettoOpenLaboratoryLimits2012,driscollWorkingBlackBox2014,grahamBigTechHarming2024,trombleWhereHaveAll2021}

		Despite these challenges, several methodological strategies can help mitigate the risks associated with the potentially incomplete and unstable datasets collected through the search endpoint. First, researchers may opt for an entirely different data collection method. Approaches based on crawling \parencite{riederMappingYouTube2020},  splitting queries in subtopics and using AND operators to increase precision (\parencite{efstratiou_youtube_2025}),
		or random sampling \parencite{zhouCountingYouTubeVideos2011,mcgradyDialingVideosRandom2023}, may be attractive for larger studies. Researchers also have the option to use channels, which represent less of a data collection problem, as an entry point.
		
		Second, especially researchers analyzing contemporary events should initiate data collection as early as possible to avoid losing access to relevant content. While questions about what we have called the `head'---the moving 20-day period yielding the highest number of results---remain, particularly for large volume queries, searches falling within this window will produce the best results.
		
		Third, repeating searches can help maximize dataset coverage and account for variations in query results. This can mean making a search per day over a longer timeframe but even running one search after another can yield previously undiscovered videos. In a small experiment using the high-volume query `chatgpt' and focusing on a single day, we were able to collect 773 videos when searching ten times in a row instead of the initial 456 results.
		
		Fourth, using several different keywords or keyword combinations may yield a larger set of videos and compensate for the opaque matching behavior we observed. This may, however, increase the complexity of subsequent filtering stages.
		
		Most of these strategies come with additional costs, both in terms of time and API quota units, and researchers will need to decide whether these investments are justified and feasible within the context of their specific research designs. While we generally recommend a combination of exhaustive data collection and purposeful filtering, whether automated or manual, some level of uncertainty remains in the current situation. Regardless of specific methodological choices, we hope that researchers using YouTube’s API will become more cognizant of these inherent limitations and document their decisions more transparently, facilitating validity assessments. Although the repeatability of data collection is heavily compromised in any case, knowledge of the exact timeframes and parameters used would provide readers with a clearer understanding of the reliability and generalizability of a given study.
		
	\section{Conclusion}
		This paper has documented significant problems regarding query matching, completeness, temporal distance, and consistency in the search endpoint of YouTube’s Data API. Most importantly, researchers hoping to study issues or events lying more than 60 days in the past risk collecting highly incomplete samples, with many or most relevant videos missing. The quota increase through YouTube’s academic research program is appreciated, but the assertion that it grants ‘access to global video metadata across the entire public YouTube corpus’ \parencite{YouTubeResearchHow} is only valid in theory given the observed behavior and may foster an undue sense of confidence. 
		
		Because the Data API’s search functionality is currently inadequate for meeting the DSA’s requirements under Article 40(4), researchers are unable to effectively support ‘the detection, identification and understanding of systemic risks in the Union,’ which include the dissemination of illegal content, negative effects on fundamental rights, and the integrity of electoral processes, or to assess ‘the adequacy, efficiency and impacts of the risk mitigation measures.’ Without the ability to reliably locate relevant content in the first place, the foundational premise of independent research collapses: how can researchers evaluate risks posed by content they cannot find?

		While this paper can hopefully raise awareness among researchers about the issues at hand and offer several mitigation strategies, only YouTube can implement structural solutions. We propose two key improvements. First, YouTube should enhance the Search Endpoint. For a company of YouTube’s scale, including all publicly listed videos in the search endpoint should be feasible. Although certain decisions may be driven by competitive advantage, we do not see this as applicable in this context. As an alternative, YouTube could create a search endpoint exclusively for vetted researchers. Second, YouTube should improve documentation. While we have speculated about reasons for the observed behavior of the search endpoint in this paper, our evidence often only highlights problems or irregularities without providing strong claims about universally applicable principles. Although search and ranking systems may be probabilistic and not fully explainable in causal terms, YouTube’s engineers should be able to describe, for instance, the differences between order options more effectively than we can and offer actionable advice to researchers. Given that YouTube has been more accommodating to third-party research than most other social media platforms, we are optimistic that the company will consider these suggestions.
		
		The issues documented in this study highlight broader concerns about the opacity of large online platforms and the challenges researchers encounter when studying them. While we hold that these companies must provide better tools and guidance to the academic community, we also believe that researchers need to invest more effort into critically evaluating data access provisions. We therefore hope this paper will inspire other groups to explore the numerous remaining questions about the possibilities and limitations of studying YouTube through data collection.

\newpage			
\printbibliography

\end{document}